# PASSIVITY ANALYSIS OF DISCRETE-TIME COUNTERPARTS OF THE BILATERAL CONTROLLED TELEOPERATION SYSTEMS

[1]A.A. GHAVIFEKR, [2]M.A. BADAMCHIZADEH, [3]A.R. GHIASI, [4]F. HASHEMZADEH

[1,2,3,4]Faculty of Electrical and Computer Engineering, University of Tabriz, Tabriz, Iran
Email: [1,2,3,4]{aa.ghavifekr, mbadamchi, agiasi, hashemzadeh}@tabrizu.ac.ir

**Abstract -** This paper presents an analysis of the passivity condition for discrete-time bilateral teleoperation systems. Considering discrete-time controllers for a master-slave teleoperation system can simplify its implementation. Varieties of control schemes have been utilized for the position error based architecture of bilateral teleoperation systems and major concerns such as passivity and transparency have been studied. This paper takes into account the passivity conditions for the discrete counterparts of P-like, PD-like, and PD-like + dissipation controllers. These conditions impose bounds on the controller gains, the damping of the master and slave robots, and the sampling time which help researchers to provide guidelines to have better transparency and passive teleoperation systems. Simulation results are performed to show the effectiveness of the proposed criteria.

**Keywords -** Passivity, Teleoperation Systems, Transparency, Sampled-data Control, Bilateral Control.

## I. INTRODUCTION

Teleoperation systems include manipulation at different scales and are vital in the remote and hazardous operations such as undersea or space explorations [1], and in delicate applications such as micro-assembly and minimally invasive surgeries [2]. In [3], a throughout review of developments and theories in teleoperation systems is presented. In a bilateral teleoperation system, the first goal is that the slave robot tracks the position of the master robot, and the second goal is that the interaction force between the environment and the slave robot is accurately transformed to the master and operator. Satisfying these goals leads to have a transparent teleoperation system, meaning that through the master robot, the human operator should feel as if he/she could able to manipulate the remote environment directly. Several control approaches have been utilized to address stability issue in the presence of time delay [4]. One of the most prevalent methods is the passivity theorem. Based on this theory and scattering matrix, the stability analysis of continuous-time teleoperation systems are extensively studied [5]. Using wave variable method [6] and adaptive controllers [7] are other continuous-time schemes that have been proposed to preserve the stability of the system. These studies have been generalized to teleoperation systems with multi master/single slave robots in [8]. Although the extensive amount of studies exist for continuous-time teleoperation systems, only a few papers have been presented on the passivity or stability analysis and controller design for the discrete-time bilateral forms. Obviously, in the nowadays digitalized world, it is of both theoretical significance and practical importance to realize how a discrete-time control signal would influence the behavior of a continuous-time dynamical network. One of the most important challenges in the sampled-data area is that the passivity of a teleoperation system is not guaranteed due to energy leaks, which caused by zero order hold devices (ZOHs) [9]. Recently, researchers mainly focus on finding conditions to preserve passivity and absolute stability which are two important indices to evaluate the performance of the teleoperation systems. Passivity of the delay-free sampled-data teleoperation systems has been studied in [14]. In [15], the exact models of the ZOH and the ideal sampler are taken into account, and the absolute stability of the system is proved using the small gain theorem. In [16], using discrete-time approximate models of the master and slave robots, a general approach for robust sampled-data stabilization of teleoperation systems is presented. The sampling effect in the transparency of the bilateral teleoperation system has been taken into account in [17]. Recent studies try to get rid of the passivity conditions of the operator or the environment. Discrete-time circle criterion has been used for the absolute stability of a sampled-data haptic interaction [18].

## II. MODELING OF THE LINEAR TELEOPERATION SYSTEM

In a bilateral teleoperation system, the human operator sends position signals to the slave robot through the master robot and the master receives the position or force signals on the slave side. The general schematic of this system is illustrated in Fig. 1, where both robots and the communication channel are lumped into a linear time invariant two-port network block. It is assumed that the dynamic structures of the master and slave robots are same, with identical or discrepant parameters.

In Fig.1, $F_h^*, F_e^*$ are the operator's and the environment's exogenous input forces, respectively. The interaction between operator/master and slave/environment are denoted by $F_h$ and $F_e$, respectively. The master and the slave position and force signals are represented by $q_m(t), q_s(t), f_m(t), f_s(t)$ respectively. Also, the dynamic characteristic of the master robot, slave robot, operator and the environment are denoted by $Z_m(s), Z_s(s), Z_h(s), Z_e(s)$.

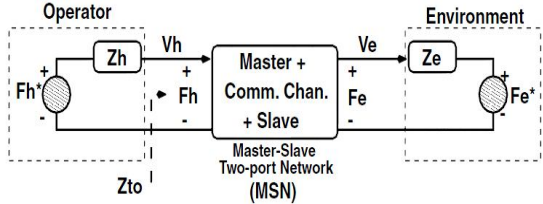

**Fig. 1. Schematic of the bilateral teleoperation network block diagram[22]**

Based on Fig.1, the dynamics of the master and slave robots can be formulated as:
$$M_m \ddot{q}_m + B_m \dot{q}_m = F_h - F_m$$
$$M_s \ddot{q}_s + B_s \dot{q}_s = F_e - F_s \quad (1)$$

The notations m and s are used for the master and slave robots, respectively. Likewise, M and B present the mass and the corresponding damping of robots. $F_m$ and $F_s$ are control inputs for the master and slave robots.

The teleoperation system in Fig.1 can be modeled as a two-port network with the following hybrid matrix presentation.
$$\begin{bmatrix} F_h(s) \\ -sq_s(s) \end{bmatrix} = H(s) \begin{bmatrix} sq_m(s) \\ F_e(s) \end{bmatrix} \quad (2)$$

where
$$H(s) = \begin{bmatrix} Z_m + C_m \dfrac{Z_s}{Z_s + C_s} & \dfrac{C_m}{Z_s + C_s} \\ -\dfrac{C_s}{Z_s + C_s} & \dfrac{1}{Z_s + C_s} \end{bmatrix} \quad (3)$$

where $Z_m = \dfrac{1}{M_m s + B_m}$, and $Z_s = \dfrac{1}{M_s s + B_s}$. $C_m$ and $C_s$ are local controllers of the master and slave robots, respectively.

For network based teleoperation systems, the scattering matrix is defined to relate the input and output wave variables by the following formula:
$$\begin{bmatrix} b_1 \\ b_2 \end{bmatrix} = \begin{bmatrix} S_{11} & S_{12} \\ S_{21} & S_{22} \end{bmatrix} \begin{bmatrix} a_1 \\ a_2 \end{bmatrix} = S(s) \begin{bmatrix} a_1 \\ a_2 \end{bmatrix} \quad (4)$$

where $[a_1 \; a_2]^T$ are input wave variables, and $[b_1 \; b_2]^T$ are output wave variables defined as following:
$$a_1 = \dfrac{F_h + R_0 V_h}{2\sqrt{R_0}}, a_2 = \dfrac{F_e - R_0 V_e}{2\sqrt{R_0}}$$
$$b_1 = \dfrac{F_h - R_0 V_h}{2\sqrt{R_0}}, b_2 = \dfrac{F_e + R_0 V_e}{2\sqrt{R_0}} \quad (5)$$

$R_0$ is the characteristic resistive impedance of the transmission line [5]. The scattering matrix can be defined in terms of the hybrid matrix:
$$S(s) = \begin{pmatrix} 1 & 0 \\ 0 & -1 \end{pmatrix}(H(s) - I)(H(s) + I)^{-1} \quad (6)$$

Analysis of the passivity in this paper is independent of the impedances of master and slave robots. It is just assumed that the environment and operator are passive. With passive but otherwise arbitrary $Z_h(s)$ and $Z_e(s)$ and based on the singular values of the scattering matrix, passivity conditions of the whole system can be derived.

Also, the linear model is considered in this paper and nonlinear terms such as friction and encoder quantization have been neglected. According to the literature [19], the friction plays a stabilizing role in a bilateral teleoperation system. The reason comes from this fact that Coulomb friction can dissipate the energy which is introduced by encoder quantization.

### III. PASSIVITY OF THE DISCRETE-TIME CONTROLLED TELEOPERATION SYSTEMS

The meaning of passivity in a linear system is equivalent to have the Nyquist diagram of the system entirely in the right half plane. A similar definition can be utilised for the passivity of discrete-time systems like the continuous-time ones. A sampled-data system is said to be passive if its power $P(k)$ consumed up to time nT satisfies following equation:
$$\sum_{k=0}^{n} P(k) = \sum_{k=0}^{n} x(k)^T y(k) \geq -\gamma$$
(7)

The system is passive if the singular value of the scattering matrix is lower than or equal to 1.

Passivity problem for the sampled-data form of the virtual wall is considered in [20]. Passivity condition is calculated as $b > KT/2 + B$, where T is the sampling time, K and B are the stiffness and damping of the virtual wall, and b is the damping of the haptic interface.

In [14], sufficient passivity conditions for the sampled-data teleoperation systems with position error based architecture have been derived. These studies impose bounds on the controller gains, the damping of the master and slave robots, and the sampling time. The proposed architecture is illustrated in Fig 2. $\tilde{f}_h$ and $\tilde{f}_e$ denote external input forces from the human operator and the environment, respectively. Two ideal samplers and two zero order holds are implied in this structure. $\alpha$ is a position scaling factor.

The following formula is used to obtain the mathematical counterpart of the sampled signal [21]:

$$X^*(s) = L\{x^*(t)\} = \sum_{k=0}^{\infty} x(kT)e^{-skT} \qquad (8)$$

By imposing the transfer function of the zero order hold to the controller inputs, they can be rewritten as:

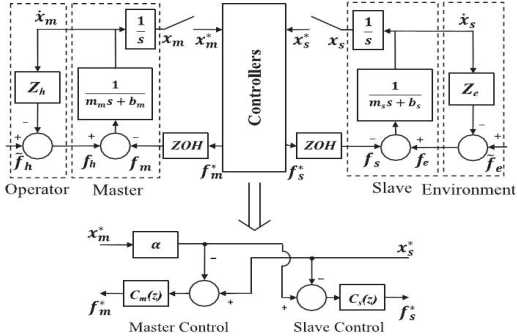

**Fig. 2. Architecture of PEB teleoperation system with discrete-time controllers and position scaling factor[17]**

$$F_m(s) = \frac{1-e^{-sT}}{sT} F_m^*(s)$$

$$F_s(s) = \frac{1-e^{-sT}}{sT} F_s^*(s) \qquad (9)$$

where, the subscript * shows sampled-time signals.
By defining $e = x_m - x_s$ as a system error, and $\alpha = 1$ as a scaling factor, the discrete-time controllers can be stated as follows:

$$F_m^*(s) = C_m(z)\big|_{z=e^{sT}} [X_s^*(s) - X_m^*(s)]$$
$$F_s^*(s) = C_s(z)\big|_{z=e^{sT}} [X_m^*(s) - X_s^*(s)] \qquad (10)$$

Using the Fourier transforms, the Parseval's theorem [22], and four constructive lemmas that are presented in [14], sufficient conditions for the passivity of a teleoperation system with discrete-time controllers are achieved as follows:

$$B_m > \frac{T}{1-\cos\omega T} \mathrm{Re}\{(1-e^{-j\omega T})C_m(e^{j\omega T})\}$$
$$B_s > \frac{T}{1-\cos\omega T} \mathrm{Re}\{(1-e^{-j\omega T})C_s(e^{j\omega T})\} \qquad (11)$$

For $\alpha \neq 1$, as a position scaling factor, the relation between controllers is stated as
$$C_m(j\omega)/\alpha = C_s(j\omega) = C(j\omega) \qquad (12)$$

In this form, the passivity conditions can be stated as[17]:

$$B_m > \frac{T(\alpha+1)}{2(1-\cos\omega T)} \mathrm{Re}\{(1-e^{-j\omega T})C_m(e^{j\omega T})\}$$
$$B_s > \frac{T(\alpha+1)}{2(1-\cos\omega T)} \mathrm{Re}\{(1-e^{-j\omega T})C_s(e^{j\omega T})\} \qquad (13)$$

## IV. CALCULATING PASSIVITY CONDITIONS FOR THREE WELL-KNOWN CONTINUOUS-TIME CONTROLLERS

This section presents the passivity conditions of discrete-time bilateral teleoperation systems. Three controllers which previously have been used for the continuous-time teleoperation systems are introduced and the passivity conditions are obtained for their discrete-time counterparts.

### 4.1 P-like Controller
The force signals which are applied by this controller on the master and slave robots are proportional of position error plus a damping injection term. The formalism of controllers is:

$$\tau_1 = K_m[q_s(t-T_1(t)) - q_m] - L_m \dot{q}_m$$
$$\tau_2 = K_s[q_s - q_m(t-T_s(t))] - L_s \dot{q}_s \qquad (14)$$

With $K_m, K_s, L_m, L_s \in R^+$. It is proofed in [28] that if the control gains satisfy the following relation:
$$4L_s L_m > (T_1^2 + T_2^2) K_m K_s \qquad (15)$$
velocity and position errors will be bounded. Also, in the case that the operator does not move the master robot and the slave robot is in free motion, the position tracking error converges to zero. By substituting the controllers (14) in the passivity condition of (13) for a delay free teleoperation system by using the bilinear transformation method we have:

$$\tau_1(z) = K_m - L_m \frac{z-1}{Tz}$$
$$\tau_2(z) = K_s - L_s \frac{z-1}{Tz} \qquad (16)$$

Thus, the passivity conditions can be simplified to:
$$B_m > K_m T + 2L_m, \quad B_s > K_s T + 2L_s \qquad (17)$$

### 4.2 PD-like Controller
The continuous-time PD-like control laws are presented as:

$$\tau_1 = K_d[\gamma_s \dot{q}_s(t-T_1) - \dot{q}_m] + K_m[q_s(t-T_2) - q_s]$$
$$\tau_2 = K_d[\gamma_m \dot{q}_s(t-T_1) - \dot{q}_s] + K_s[q_m(t-T_2) - q_s] \qquad (18)$$

Which all gains of $K_d, K_m, K_s$ are positive constants and
$$\gamma_i^2 = 1 - \dot{T}_i^2 \qquad (19)$$
$\gamma_i$ is a time-varying gain which compensate the energy generated due to the increase or decrease of the time delay. Imposing the control laws of (18) to teleoperation system, and choosing the control gains as $K_s \geq K_m$ which also satisfied (15), it is proofed in [28] that velocities and position tracking errors will be bounded.

By substituting the controllers (18) in the passivity condition of (13) and using bilinear transformation method we have:

$$\tau_1(z) = K_d \gamma_s \frac{z-1}{Tz} + K_m$$
$$\tau_2(z) = K_d \gamma_m \frac{z-1}{Tz} + K_s \qquad (20)$$

Thus, the passivity conditions can be simplified to:
$$B_m > K_m T + 2K_d \gamma_s, \quad B_s > K_s T + 2K_d \gamma_m \qquad (21)$$

### 4.3 PD-like Controller + Dissipation
This controller is proposed in [29] in order to achieve bilateral force reflection, master-slave coordination and energetic passivity of the closed loop teleoperation system. The controllers are designed as:

$$\tau_1(t) = -K_v(\dot{q}_1(t) - \dot{q}_2(t-\tau_2)) - (K_d + P_\varepsilon)\dot{q}_1(t) - K_P(q_1(t) - q_2(t-\tau_2))$$

$$\tau_2(t) = -K_v(\dot{q}_2(t) - \dot{q}_1(t-\tau_1)) - (K_d + P_\varepsilon)\dot{q}_2(t) - K_P(q_2(t) - q_1(t-\tau_1))$$
(22)

Where $T_1, T_2 \geq 0$ are delays from master to slave and vice versa. $K_v, K_p$ are the symmetric and positive definite gains, $K_d$ is the positive dissipation gain and $P_\varepsilon$ is an extra damping to protect master-slave coordination. It is proofed in [29] that choosing $K_d = \frac{v}{2} K_p$, where $v > 0$ is an upper bound of the general delay $T_1 + T_2$, leads to have a passive teleoperation system. Also, if the human operator and the environment are passive, then the tracking error signal between the master and slave will be bounded. Furthermore, if the velocity and accelerations signals converge to the zero, force tracking error will be achieved. By substituting the controllers (22) in the passivity condition of (13) and using bilinear transformation method we have:

$$\tau_1(z) = \tau_2(z) = -K_v \frac{z-1}{Tz} - (K_d + P_\varepsilon)\frac{z-1}{Tz} - K_P \quad (23)$$

Thus, the passivity condition can be simplified to:

$$B_m, B_s > K_p T + 2K_d - 2P_\varepsilon - 2K_v \quad (24)$$

## V. SIMULATION STUDY

The teleoperation system in Fig 2, has been considered, and the aforementioned passivity conditions for three continuous-time controllers have been tested for 1-DOF master and slave robots modelled by the mass and damping terms. By simulation, we can study the effect of sampling rate on the behavior of discrete-time teleoperation system for different architectures of controllers. The simulation parameters for the P-like controller are chosen as $K_m = K_s = 1$ and $L_m = L_s = 0.1$. For the PD-like controller the parameters are $K_d = 1$ and $K_m = K_s = 2$. As we consider the delay-free teleoperation system then $\gamma_i = 1$. The chosen parameters for the PD-like controller with dissipation are $K_p = 1, K_d = 2, P_\varepsilon = 0.002, K_v = 10$. The environment acts like a wall, thus reflecting the entire torque of the slave. We consider the following scenario. The human operator pushes a step force to the master robot for 5 seconds from 10s till 20s. The contact with the environment occurs when the slave robot reaches $4 rad$. Human operator assumed to behave such as PD tracking controller with its spring and damping gains as $10 N/m$ and $1 Ns/m$. The implied force by the operator is depicted in Fig.3. Both master and slave robots are a linear one degree of freedom system with transform function $M(s) = 2.2/(4 + 3.5s)$. Figures 4 to 6 present the position signals of master and slave robots by choosing 0.002s for P-like, 0.005s for PD-like, 0.006s for PD-like+ dissipation as maximum allowed sampling times, respectively. Also, force tracking signals for the PD-like controller +dissipation are depicted in Fig 7. The most noticeable result is that increasing controller gains and sampling time jeopardizes the passivity condition. On the other hand, the physical characteristic of the robot cannot be changed easily. Besides passivity, transparency is another important factor in teleoperation systems. As mentioned in section 2, increasing the controller gains leads to the more transparent system in the continuous-time teleoperation systems. But, in the discrete-time structures, there should be a trade-off between the passivity conditions of (13) and transparency of the system.

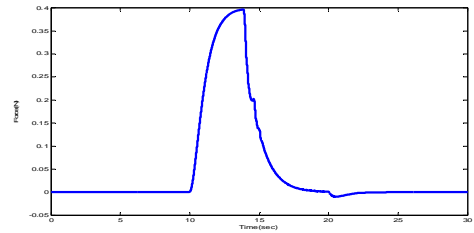

**Fig. 3. External force applied by the operator**

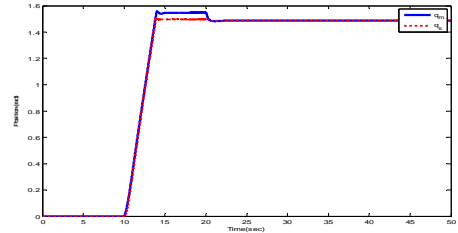

**Fig. 4. Master and slave robots position signals for discrete-time counterpart of P-like controller**

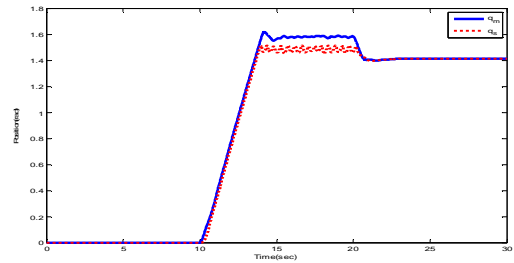

**Fig. 5. Master and slave robots position signals for discrete-time counterpart of PD-like controller**

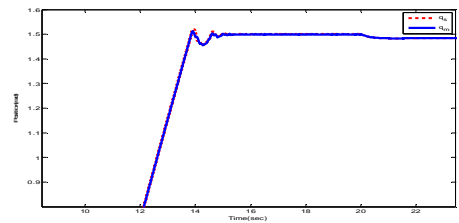

**Fig. 6. Master and slave robots position signals for discrete-time counterpart of PD-like+ dissipation controller**

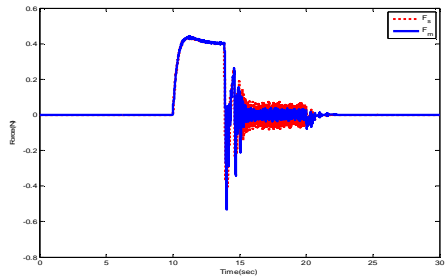

**Fig. 7. Master and slave robots force signals for discrete-time counterpart of PD-like+dissipation controller**

## CONCLUSION

Despite the extensive amount of research concerning continuous-time teleoperation systems, only a few papers have been published on the analysis and controller design for discrete bilateral forms. This paper evaluated discrete-time controllers for teleoperation systems, including issues such as transparency and passivity. Passivity conditions have been found for the discrete-time counterpart of three well-known controllers including P-like controller, PD-like controller and PD-like+ dissipation controller. For this purpose, models of the ZOH and the sampler are incorporated in an appropriate frequency-domain analysis. The aforementioned conditions impose a lower bound on the robot damping and an upper bound on the sampling time. It is concluded from the simulation part that by using the PD-like+ dissipation controller we can choose higher sampling time without losing passivity.

## REFERENCES

[1] E. F. Cardenas and M. S. Dutra, "An Augmented Reality Application to Assist Teleoperation of Underwater Manipulators," IEEE Latin America Trans., vol. 14, pp. 863-869, 2016.

[2] A. A. Ghavifekr, M. Badamchizadeh, G. Alizadeh, and A. Arjmandi, "Designing inverse dynamic controller with integral action for motion planning of surgical robot in the presence of bounded disturbances," in Electrical Engineering (ICEE), 2013 21st Iranian Conference on, 2013, pp. 1-6.

[3] P. F. Hokayem and M. W. Spong, "Bilateral teleoperation: An historical survey," Automatica, vol. 42, pp. 2035-2057, 2006.

[4] H. Du, "H∞ state-feedback control of bilateral teleoperation systems with asymmetric time-varying delays," IET Control Theory & Applications, vol. 7, pp. 594-605, 2013.

[5] H.-C. Hu and Y.-C. Liu, "Passivity-based control framework for task-space bilateral teleoperation with parametric uncertainty over unreliable networks," ISA transactions, 2017.

[6] D. Sun, F. Naghdy, and H. Du, "Application of wave-variable control to bilateral teleoperation systems: A survey," Annual Reviews in Control, vol. 38, pp. 12-31, 2014.

[7] Z. Chen, Y.-J. Pan, and J. Gu, "Adaptive robust control of bilateral teleoperation systems with unmeasurable environmental force and arbitrary time delays," IET Control Theory & Applications, vol. 8, pp. 1456-1464, 2014.

[8] J. Li, M. Tavakoli, and Q. Huang, "Stability of cooperative teleoperation using haptic devices with complementary degrees of freedom," IET Control Theory & Applications, vol. 8, pp. 1062-1070, 2014.

[9] M. Tavakoli, A. Aziminejad, R. Patel, and M. Moallem, "Discrete-time bilateral teleoperation: modelling and stability analysis," IET Control Theory & Applications, vol. 2, pp. 496-512, 2008.

[10] A. Jazayeri and M. Tavakoli, "A passivity criterion for sampled-data bilateral teleoperation systems," Haptics, IEEE Trans. on, vol. 6, pp. 363-369, 2013.

[11] A. Jazayeri and M. Tavakoli, "Absolute stability analysis of sampled-data scaled bilateral teleoperation systems," Control Engineering Practice, vol. 21, pp. 1053-1064, 2013.

[12] H. Beikzadeh and H. J. Marquez, "Robust Sampled-Data Bilateral Teleoperation: Single-Rate and Multirate Stabilization," IEEE Trans. on Control of Network Systems, 2016.

[13] T. Yang, Y. L. Fu, and M. Tavakoi, "An analysis of sampling effect on bilateral teleoperation system transparency," in Control Conference, 2015 34th Chinese, 2015, pp. 5896-5900.

[14] N. Miandashti and M. Tavakoli, "Stability of sampled-data, delayed haptic interaction under passive or active operator," IET Control Theory & Applications, vol. 8, pp. 1769-1780, 2014.

[15] G. Leung, "Bilateral control of teleoperators with time delay through a digital communication channel," in Proc. 30th Annual Allerton Conf. on Communication, Control, and Computing, 1992, pp. 692-701.

[16] J. E. Colgate and G. G. Schenkel, "Passivity of a class of sampled-data systems: Application to haptic interfaces," Journal of robotic systems, vol. 14, pp. 37-47, 1997.

[17] K. Ogata, Discrete-time control systems vol. 2: Prentice Hall Englewood Cliffs, NJ, 1995.

[18] A. Oppenheim and A. Willsky, "with SH Nawab, Signals and Systems," Prentice—Hall, vol. 1, p. 997, 1997.

[19] E. Nuño, L. Basañez, R. Ortega, and M. W. Spong, "Position tracking for non-linear teleoperators with variable time delay," The International Journal of Robotics Research, vol. 28, pp. 895-910, 2009.

[20] D. Lee and M. W. Spong, "Passive bilateral teleoperation with constant time delay," IEEE transactions on robotics, vol. 22, pp. 269-281, 2006.

★ ★ ★